# Advancing Accessible Hand-Arm Vibration Safety Monitoring: ISO-Compliance with Wearable Sensors and Transfer Functions

Johannes Mootz[1] and Reza Akhavian, Ph.D.[2]

[1]Data-informed Construction Engineering (DiCE) Research Lab, Department of Civil and Environmental Engineering, San Diego State University. E-mail: jmootz0676@sdsu.edu
[2]Associate Professor, Data-informed Construction Engineering (DiCE) Research Lab, Department of Civil and Environmental Engineering, San Diego State University (corresponding author). E-mail: rakhavian@sdsu.edu

## ABSTRACT

Field workers are frequently exposed to hazardous vibrations, increasing the risk of Hand-Arm Vibration Syndrome (HAVS) and other long-term health problems. ISO 5349-1 provides guidelines for measuring vibration exposure. However, this standard was established in controlled conditions using high-quality accelerometers directly attached to power tool handles. This study investigates an alternative, wearable sensor-based data collection process and develops an error-minimization transfer function that derives values comparable to ISO benchmarks for safety monitoring. Experiments are performed with subjects hammer drilling into concrete while vibrations are measured using three accelerometers at different sampling frequencies. The transfer function maps vibration data across sensor positions by accounting for damping effects. The findings indicate a significant reduction in acceleration between the palm and upper arm, highlight the impact of sampling frequency on data accuracy, and enable accurate comparison of true hand-arm vibration levels with existing standard limits to allow accessible, real-time, and cost-effective HAVS prevention.

## INTRODUCTION

Humans encounter various vibration sources in daily life, with exposure levels depending on the work environment and activities. In industries such as construction, transportation, agriculture, and manufacturing, hand-transmitted vibrations (HTV) are common among workers operating vibrating tools and machinery (Aizuddin and Jalil 2022). Prolonged exposure to these tools can lead to hand-arm vibration syndrome (HAVS), a condition characterized by vascular, musculoskeletal, and neurological symptoms, including tingling, numbness, and episodic finger blanching triggered by cold stimuli, known as vibration white finger (Griffin 2012; Zimmerman et al. 2017). The National Institute for Occupational Safety and Health (NIOSH) in the U.S. has conducted extensive research on HTV risks, significantly advancing knowledge in this area (Dong et al. 2021).

  The effects of tool vibrations on workers' joints are shaped not only by their intensity but also by their frequency, underscoring the importance of precise measurement and control (Reynolds and Angevine 1977). To manage the risk of hand-arm vibration (HAV) exposure, the standard ISO 5349-1 (ISO 2001a) establishes guidelines for evaluating and controlling daily exposure levels, requiring employers to adhere to legally defined limits. ISO 5349-1 and 5349-2 (ISO 2001b) outline exposure measurement methods, emphasizing vibration magnitude and duration. However, compliance requires tool-mounted accelerometers, which may limit routine data collection due to coupling effects (Maeda et al. 2019; 2020).

Previous research has mostly focused on the vibration source of tools, aiming to optimize their design, reduce the intensity of hand-transmitted vibrations, and minimize their detrimental effects on the human hand-arm system. The transmissibility provides information regarding the rates of vibration at different locations within the hand-arm system. (Adewusi et al. 2010). Vibration transmission characteristics are affected by factors such as hand-arm posture, vibration frequency and amplitude, and grip force (Zhang et al. 2021).

It is of high importance to leverage innovative technologies to bridge the gap between understanding these effects and accurately measuring them in real-world scenarios. For example, implementing wearables can be an effective approach to quantify these vibrations. Specifically, analyzing data from inertial measurement unit (IMU) sensors, either worn on the user's body (Casale et al. 2011) or integrated into the user's smartphone carried by the user (Kwapisz et al. 2011), provides an efficient method for tracking movements. In a construction context, smartphones and their embedded accelerometers have been used for activity recognition applications (Akhavian and Behzadan 2016). In general, wearable sensor technology can help detect potential risks to the human body, to prevent bodily harm (Maeda et al. 2020). This offers several significant advantages, including the capacity for real-time monitoring. This capability is instrumental in ensuring the safety of construction workers by promptly identifying any instances where the ISO standard is not being met.

Real-time vibration monitoring using accelerometers in construction remains understudied, with most research focusing on Ecological Momentary Assessment (EMA). EMA continuously tracks time-varying variables in real-world settings using portable devices, reducing recall bias in self-reports reports (Ramsey et al. 2016; Seong et al. 2022). EMA collects data on symptoms, emotions, and behaviors via portable devices like smartphones and actigraphy (Engelen et al. 2016). However, accurately measuring vibrations requires accelerometers, which detect subtle vibrations often missed in construction settings (ISO 2001a).

This study aims to develop a cost-effective yet reliable system to monitor vibrations in real-time, using a wearable device or smartphone approach for accessible and accurate data collection and further processing. Towards this goal, three areas of uncertainty are explored here: (1) the difference between the measured vibration at the handle and the power tool holding hand, (2) the transmission of vibration through the non-linear properties of the hand-arm system, and (3) the difference between the spectrum of the measured vibration at the upper arm and the power tool holding hand at a lower sampling frequency.

**METHODOLOGY**

Experiments were conducted to measure the vibration at different positions in the hand-arm system. Here, three 3-axis, customer-grade, cost-effective, MEMS-based accelerometers namely, HWT906-TTL MPU-9250, WT9011DCL MPU9250 (WitMotion Shenzhen Co., Ltd), and Noraxon Ultium Motion (NUM) model 870 (Noraxon U.S.A. Inc., Scottsdale, AZ, USA) were deployed. The three accelerometers used in this study were selected based on accuracy, sensitivity, and durability to ensure reliable vibration measurements. Cost-effectiveness and compatibility with the wireless smartphone-based data acquisition system were considered to allow for practical implementation in the field. Table 1 displays an overview of the key characteristics of these accelerometers. While ISO 5349-1 specifies up to 1250Hz, this study prioritized accessible, cost-effective sensors, accepting this trade-off for practicality.

**Table 1. Overview of utilized accelerometers.**

|  | **HWT906-TTL MPU-9250** | **WT9011DCL MPU9250** | **Noraxon Ultium Motion (NUM) 870** |
|---|---|---|---|
| Frequency range [Hz] | 0.2-1000 | 0.2-200 | 0-400 |
| Measuring range | ±16g | ±16g | ±200g |
| Wireless | No | Yes (Bluetooth 5.0) | Yes |

As illustrated in Figure 1, the HWT906 was attached to the handle of the power tool (and it will be referred to as the power tool sensor hereinafter), with a customized connecting interface. As illustrated in Figure 2, the NUM sensors were attached at six different body positions (RT=right, LT=left): HandRT, HandLT, ForearmRT, ForearmLT, UpperArmRT, and UpperArmLT. The WT9011 sensor was placed directly next to the UpperArmRT sensor (and it will be referred to as the near UpperArmRT sensor hereinafter). This allows it to be compared with the more reliable NUM system and provides insight into the measurement of the same vibration with different frequency ranges.

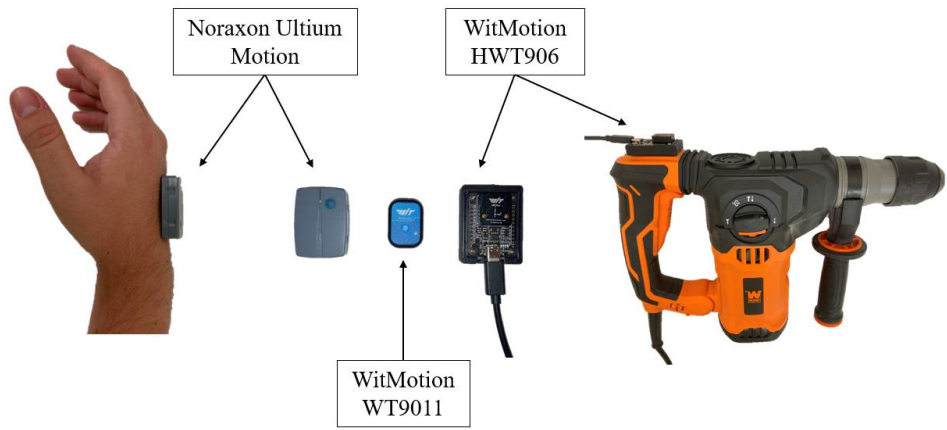

**Figure 1. Size comparison of the wearables utilized in this study.**

Four subjects performed hammer drilling into concrete. Given the focus on achieving mechanistic insights rather than statistical generalization and inference across a population, this sample size was sufficient to demonstrate the fundamental physical relationships being investigated. The experiments were conducted using the rotary hammer model RH1042 (WEN Products, Inc.), a tool representative of widely used rotary hammers in construction applications, particularly for tasks involving high-intensity vibration exposure. This model was selected based on its availability, reliability, and compliance with industry. The experimental design comprised two runs, with each run lasting one minute and encompassing the use of two different tools: a 5/16'' drill bit and a 9/19'' flat chisel. Prior to each trial, all sensors were systematically calibrated. All subjects used their dominant right hand on the rear handle of the rotary hammer, where the power tool sensor was located. Several external factors, such as grip strength variations, operator fatigue, and environmental vibrations, have the potential to influence vibration readings. To mitigate fatigue effects, experiments were conducted in short intervals, while environmental vibrations were minimized by ensuring all drilling occurred in the same concrete substrate. In contrast, grip force was left to the discretion of each participant to reflect natural variations in tool handling, aligning with real-world working conditions.

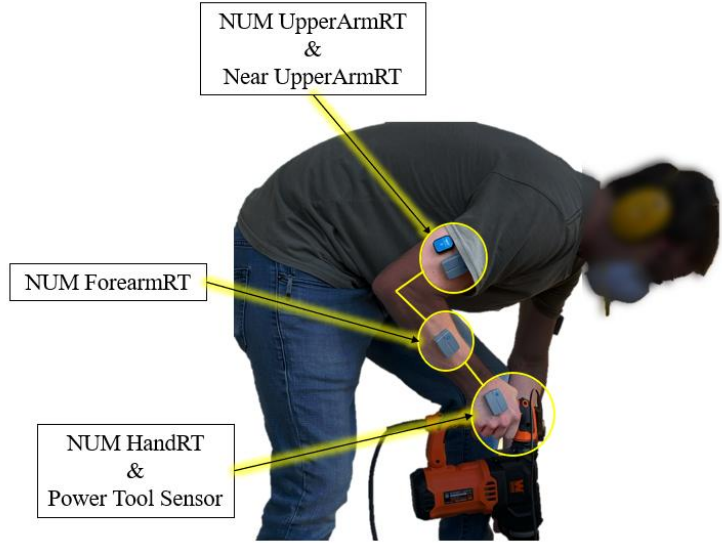

**Figure 2. Sensor placement on the subject's right arm.**

MATLAB, Version 2023b (The MathWorks, Inc., Natick, MA, USA) was used for data analysis. The data was divided into 10-second segments, yielding a total of 50 usable time series, each comprising 4000 data points. The data obtained from the accelerometer includes not only the measured acceleration but also the component of gravity. Given the dynamic nature of the experiment, it is not possible to guarantee that the sensor axes will be aligned at all times. Consequently, the influence of gravity is eliminated through the implementation of a low-pass filter, which captures the fundamental motion of the subject and subsequently subtracts it from the raw data, as shown in Equation 1.

$$a_{dyn}(i) = a(i) - g_{est}(i) = a(i) - \beta \cdot a(i) + (1 - \beta) \cdot g_{est}(i - 1) \qquad (1)$$

Here, $g_{est}(i)$ is the estimation of the gravity component at sample $i$ for the instantaneous acceleration $a(i)$. A weighting parameter $\beta = 0.05$ was determined to be a suitable value based on empirical observation. This results in acceleration $a_{dyn}$ that is independent of the gravitational component. To maintain the integrity of the high-frequency components of the experiment, the implementation of de-noising was deemed unnecessary.

As prescribed in ISO 5349-1 the three-axial acceleration is used to determine a measure for vibration. The root-mean-square value (RMS) of the instantaneous acceleration values of each axis is calculated as shown in Equation 2 (only presented for the x-component). Also, frequency-weighting is applied to account for pre-determined effects of the vibration. This is summed up to the vibration total value of frequency-weighted RMS acceleration, as shown in Equation 3 presented for the x-component.

-4-

$$a_{hwx} = \sqrt{\frac{1}{N}\sum_{n=1}^{N} a_{hwx}(t)^2} \qquad (2)$$

$$a_{hv} = \sqrt{(a_{hwx}^2 + a_{hwy}^2 + a_{hwz}^2)} \qquad (3)$$

N is the total number of instantaneous acceleration values $a_{hvx}(t)$ (here presented for the x-axis). The RMS values are combined with the acceleration total value $a_{hv}$.

To achieve a more comprehensive understanding of the dynamics of the hand-arm system, a transfer function is derived by employing a Box-Jenkins model, as described in Equation 4. The Box-Jenkins model was chosen because it separately models system dynamics and noise, making it well-suited for handling vibration data with significant noise measurement. Unlike autoregressive models such as ARX or ARMAX models, Box-Jenkins establishes two distinct systems for the actual system dynamics and noise, leading to more accurate vibration analysis. Its efficacy is particularly pronounced in scenarios where noise contributes significantly and independently to the output, ensuring more accurate vibration analysis and reliable predictions. The model functions as a one-step-ahead predictor, incorporating past inputs and outputs. Here, the input is assumed as the NUM acceleration sensor at the right hand (i.e., HandRT), and the output is defined as the NUM acceleration sensor at the right upper arm (i.e., UpperArmRT).

$$y(t) = \frac{B(q)}{A(q)} u(t-1) + \frac{C(q)}{D(q)} e(t) \qquad (4)$$

$$B(q) = b_1 + b_2 q^{-1} + \cdots + b_n q^{-n} \qquad (5)$$

$$A(q) = 1 + a_1 q^{-1} + \cdots + a_n q^{-n} \qquad (6)$$

where *y(t)* denotes the output at timestep *t*. The input signal *u(t-1)* and the white noise *e(t)* serve as inputs for the model. As shown in Equations 5 and 6, *A(q), B(q), C(q)*, and *D(q)* are independent parametrizations with the shift operator *q* for this model. They allow the independent characterization of the noise. A comparison was made between the model's simulated and predicted outputs and the actual outputs (the simulated output does not consider the noise model, while the predicted output does). This allows the influence of noise to be studied. The transfer function was envisioned to demonstrate the gain at different frequencies and to provide insights into the frequency dependency.

A key property of the Box-Jenkins model is the model order, which indicates the number of *n* time shifts in Equations 5 and 6. A low model order may not capture all relevant aspects, while a high model order increases the computational cost. Therefore, the model order was systematically tested. To evaluate the model, the RSME was calculated from the RMS values of the predicted and actual values.

# RESULTS

Statistical analyses were carried out using IBM SPSS Statistics, Version 29.0.0.0 (SPSS Inc., Chicago, IL, USA) and a p-value of <0.05 was set to determine statistical significance. To determine statistically significant differences, a double-sided paired t-test was performed. This t-test was conducted for sensors that were placed at the same location, where the same measurements were anticipated (HandRT and power tool sensor; UpperArmRT and near UpperArmRT sensor) Additionally, for sensors at different positions, linear regressions were performed to examine relations between these sensor combinations (HandRT and ForearmRT; ForearmRT and UpperArmRT; HandLT and ForearmLT; ForearmLT and UpperArmLT). The RMS value serves as the metric here.

The paired t-test revealed no statistically significant difference between HandRT RMS and the power tool sensor vibration RMS values (t(18) = -0.006, p=0.995) among the participants. Note that an unreliable cable connection introduced issues that resulted in fewer samples for the power tool sensor here. However, a significant divergence was observed between the UpperArmRT RMS and the near UpperArmRT sensor vibration RMS values (t(49) = -10.519, p <0.001). The I/O relationship of the signals is displayed in Figure 3a. HandRTx denotes the input, while UpperArmRTx represents the output. Additionally, the magnitudes of the eight sensors for the 50 segments are displayed in Figure 3b.

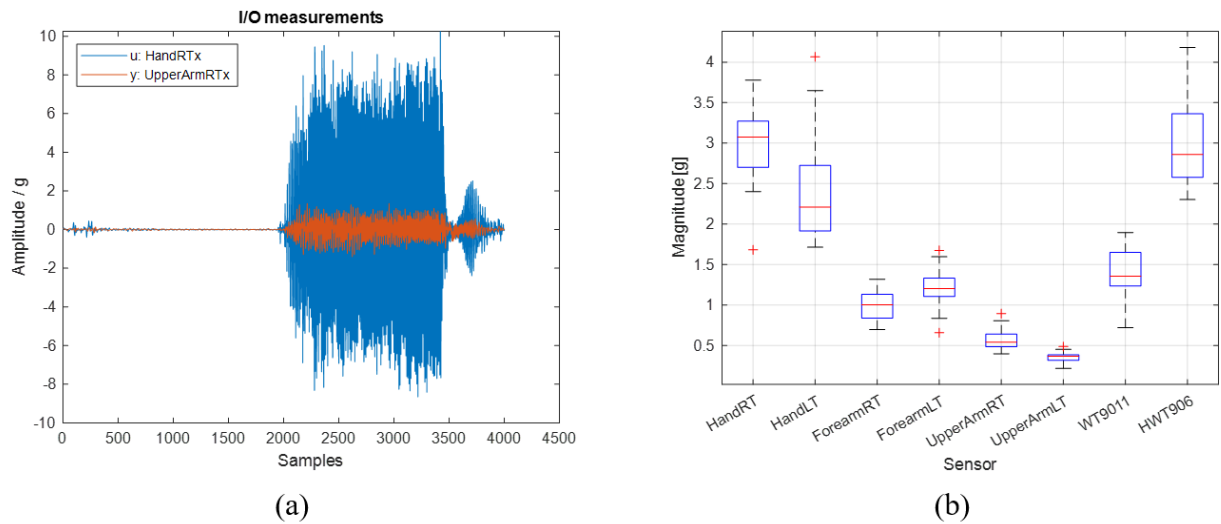

**Figure 3. (a) I/O measurements for one exemplary sample (subject 1, run1) and (b) boxplots of the RMS value of the eight accelerometers.**

Linear regression was used to test if the four different sensor combinations significantly predict each other. Table 2 shows the results of the four sensor combinations.

After systematic tests, the order of the Box-Jenkins model was determined to be 20, which provided an acceptable fit between accuracy and cost. The transfer function was evaluated for each of the 50 segments.

**Table 2. Results of the linear regression between four different sensor combinations with significant values in bold.**

|  | HandRT → ForearmRT | ForearmRT → UpperArmRT | HandLT → ForearmLT | ForearmLT → UpperArmLT |
|---|---|---|---|---|
| Adjusted $R^2$ | 0.524 | 0.0135 | 0.609 | -0.0126 |
| F(1,48) | 54.8 | 1.67 | 77.4 | 0.389 |
| p-value | **<0.0001** | 0.203 | **<0.0001** | 0.536 |

Figure 4 illustrates several aspects for one of the 50 segments: (a) the calculated transfer function consisting of a noise model *H(q)* and a system model *G(q)*, (b) a comparison of measured, simulated, and predicted values, and (c) an enlarged detail of the same figure. It is noteworthy that the noise model exhibits a higher gain compared to the system model in this instance. Additionally, the predicted output (including the noise) closely resembles the measured output, outperforming the simulated output (excluding the noise). This discrepancy is further substantiated by the RMSE values, which measure the discrepancy between the simulated and predicted outputs. The RMSE values for the simulated and predicted outputs are 0.2064 and 0.0424, respectively.

The prediction model's normalized RMSE, evaluated for each axis, yielded the following values: 85.06% for the x-axis, 81.82% for the y-axis, and 84.86% for the z-axis. The RMS values demonstrate a discrepancy of 0.88% between the observed and predicted values.

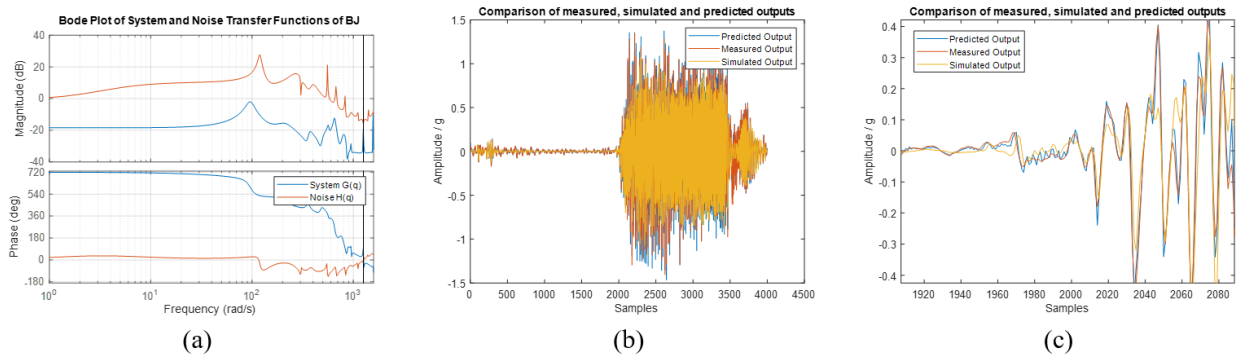

(a)          (b)          (c)

**Figure 4. (a) transfer function, (b) comparison of measured, simulated, and predicted outputs for one exemplary segment, (c) enlarged figure of (b).**

## DISCUSSION AND CONCLUSION

The t-test reveals no statistically significant difference between the HandRT and the power tool sensor vibration values. This finding suggests that, despite a different placement from what is prescribed in ISO 5349-1, the accelerometer consistently generates reliable results with a 95% confidence level. Consequently, this alternative sensor placement could be regarded as a potential solution, as it would enable direct measurement of acceleration at the user's body, thereby potentially minimizing interference with the worker's task. The potential implications of this

-7-

approach include the elimination of reliance on the handle attachment as in ISO 5349-1 and the utilization of well-established wearables.

The t-test also reveals an inability to derive reliable conclusions from the observed relationship between the UpperArmRT and the near UpperArmRT sensor vibration values. Despite the expectation that the results would be consistent, the t-test indicates a significant discrepancy. This discrepancy could be attributed to several factors, including the inadequate sampling frequency of 200Hz, which might not have captured all the relevant effects. Also, WitMotion accelerometers have an internal bandwidth filter of 0-256 Hz which cannot be disabled and therefore distorts results. However, given the observation that the RMS value of the near UpperArmRT sensor is larger than that of the UpperArmRT counterpart, this discrepancy may be attributable to factors other than the sampling frequency, such as poor attachment of the sensor.

Linear correlation analysis has been conducted to determine relationships between sensors that are not directly adjacent to each other. Results indicate the potential for predicting forearm values based on hand values for both the left and right sides. A linear correlation has been identified between these two sensor placements, underscoring the interpretative value of the accelerometer data collected from the sensor placed on the forearm. This finding suggests that the forearm could be used as a potential placement for wearable devices to determine safe vibration levels.

Conversely, it has been determined that a significant linear correlation between the forearm and the upper arm for both the left and right sides is not present. Consequently, it is not feasible to draw reliable conclusions between the forearm and the upper arm, at least using a linear regression model. This discrepancy can be explained by the elbow joint's natural capacity to modify vibration characteristics between the two measuring points. Additionally, in practical applications, users often stiffen their wrists to a greater extent than their elbows, which could be a contributing factor to the observed differences. Future work could explore the use of alternative non-linear functions or the application of tools such as principal component analysis to improve this investigation.

To understand if readings from sensors other than those attached to a vibratory device can be used to evaluate safe vibrations as prescribed by ISO standards, a transfer function was also evaluated. The discrepancy between the actual and the predicted RMS value is minimal (0.88%). This indicates that the Box-Jenkins model with a model order of 20 provides a valid representation of the transfer function between the hand and upper arm. The model's structure, incorporating separate noise and system dynamics, is justified by the dominance of the noise model in the magnitude gain. This finding highlights the difficulty in drawing reliable conclusions from hand data to upper arm data. The simulation represents only the system model while the prediction includes the system model and the noise model. While the prediction approach can achieve sufficiently accurate results, the simulation model lacks accuracy. This reveals its limitations as the large noise model amplifies white noise, which does not carry additional information about the system. Consequently, predicting upper arm values from hand values remains inherently challenging.

While this study investigates the effectiveness of simple but effective data collection to ensure worker safety when operating vibration tools, it has some limitations. The first limitation pertains to the use of low-cost customer-grade accelerometers. These instruments are incapable of ensuring an adequate range of measurable frequencies as prescribed in the ISO standards, thus hindering the capacity to generate precise results accordingly. Addressing this issue requires a more in-depth analysis with easy-to-access and cost-effective sensors or developing different standards other than those of the ISO. Additionally, while the sample size was sufficient for demonstrating the fundamental relationships investigated in this study, increasing the number and diversity of participants would enhance the generalizability of the findings across different worker populations. Furthermore, the study focused on a single task with one power tool, which, while representative of common construction activities, limits the applicability of the results to a broader range of tools and operational conditions. Since factors such as body position, grip force, feeding force, and anthropometric properties influence vibration transmission, future research could explore a wider array of tools and tasks to provide more comprehensive insights. Also, this study examined immediate vibration exposure rather than long-term effects. Incorporating extended monitoring and assessing potential health outcomes over time could further strengthen the practical relevance of this research for HAVS prevention. Lastly, while this study explores an alternative approach to measuring hand-arm vibration, a direct comparison with ISO 5349-1 measurements was not conducted in full detail. The results indicate that measuring vibration at the hand or forearm can provide reliable data, but further studies should systematically compare these values with tool-mounted measurements prescribed by ISO standards. Such an analysis would strengthen the validation of the proposed method and its applicability for standardized safety assessments.

This study evaluated the feasibility of measuring hand-arm vibration at various anatomical locations to identify alternative measurement locations beyond the tool-mounted approach prescribed by ISO 5349-1. This approach could facilitate the implementation of more suitable methods for assessing vibration in real-world applications, which differ from the controlled environment of a laboratory. The results indicate that while vibration measurement is not strictly limited to the power tool, optimal sensor placements include the hand and forearm, while the upper arm is less suitable. The transfer function derived from the Box-Jenkins model accounts for damping effects by mapping hand and forearm vibration levels to equivalent tool-based values that are directly comparable to ISO limits. This approach offers several advantages, including minimal interference with tool operation, continuous availability regardless of tool changes, and compatibility with low-cost, compact sensors and wearables. By enhancing the feasibility of real-time vibration monitoring in dynamic work environments, accessible and cost-effective HAVS prevention can be supported.